\documentclass[aps,prl,twocolumn,groupedaddress]{revtex4}
\usepackage{graphicx}
\begin{document}

\title{Kafri, Mukamel, and Peliti Reply}

{\bf Kafri, Mukamel, and Peliti Reply}: The Comment of Hanke and
Metzler \cite{HM} questions the validity of the analysis presented
in \cite{KMP1} to DNA chains of finite length as used in
experiments. Their argument is that for the analysis to be valid
``each of the three segments going out of a vertex must be much
longer than the persistence length $\ell_p$ of this segment''. By
using the persistence lengths $\ell_p(L) \sim 40$\AA~for a single
strand and $\ell_p(H) \sim 500$ \AA~for a double helix (bound
segment) they arrive at the conclusion that in order to observe
the asymptotic behavior found in \cite{KMP1} one needs chains
which are far longer than those studied experimentally.

This assertion is a consequence of a misinterpretation of the
analysis given in \cite{KMP1}. In this analysis one considers a
loop interacting with {\it the rest of the chain} and not just
with the vicinal double helices. Thus, in \cite{KMP1} each of the
two lines attached to the loop is in fact composed of an
alternating sequence of bound segments and denaturated loops. It
may be viewed as a stick-and-joint structure, whereby adjacent
double helices are loosely attached to each other via an open
loop. The ``rest of the chain'' as considered in reference
\cite{KMP1} is in fact a jointed rods structure with a persistence
length which is given by $\ell_p(H)$ when the distance between
loops, $\xi$, is larger than $\ell_p(H)$ and is given by $\xi$
when $\xi < \ell_p(H)$.

In the Comment the authors claim that the analysis of \cite{KMP1}
is valid only when the helical segment is flexible and thus much
larger than the persistence length $\ell_p(H)$, where it could be
described by a self-avoiding walk. This claim is incorrect. The
analysis is valid as long as the chain contains a sufficient
number of loops to allow considering the stick-and-joint structure
(Fig. 1) as a self-avoiding walk. Thus, even short chains of
length of the order of $5000$ base pairs could allow for about
$20$ joints which is large enough to make the analysis of
\cite{KMP1} valid. This is clearly evident from experimental
melting curves \cite{Blake1,WB} of chains of such lengths. These
curves exhibit over $10$ individual sub transitions corresponding
to the existence of a large number of loops (since some of the
peaks correspond to more than one loop). Thus, the claim made in
\cite{HM} that the Fisher exponent $c=1.766$ should be employed
when the bound segment is a rigid rod is not valid.
Moreover, many experiments were carried out on chains of a length
of the order of $10^6$ base pairs \cite{Blake2} where the
asymptotic regime considered in \cite{KMP1} is easily accessible.

Our observation \cite{KMP1} that the effective loop entropy
parameter $c$ is in fact larger than 2 affects other parameters
which have been applied in modeling DNA. One such example is the
cooperativity parameter $\sigma_0$, which affects the melting
curve rather drastically.
Within the Poland-Scheraga (PS) approach the average distance
between loops near the transition is given by $\xi \approx
1/\sigma_0$ for $c>1$. This parameter has been estimated in
various studies to be very small, of the order of $10^{-3}$ to
$10^{-5}$ \cite{Blake1,WB,Amir,Zimm,Blake2}. We point out that
this small numerical value is obtained by fitting experimentally
sharp melting curves to a theory which yields a {\it continuous
transition} \cite{Amir,Zimm} (namely, the PS model with $1<c<2$),
or sometimes {\it no transition at all} \cite{Blake2} ($c=0$).
Small $\sigma_0$ makes a continuous transition look sharp,
yielding a good fit with experiment. In particular, $\sigma_0$ is
usually estimated by considering the maximal value of the
temperature derivative of the order parameter $\theta$ near the
transition. For $c<2$ one has \cite{PS} $\Psi \equiv -\left(
\partial \theta/\partial T \right)_{\rm max} \propto
\sigma_0^{-1/(2-c)}$. The large observed $\Psi$ yields a very
small $\sigma_0$ when fit with, for example, $c=1.5$, $1.766$
\cite{Zimm,Amir}, or $c=0$ \cite{Blake2}. However, since the
effective $c$ is larger than $2$, the transition is expected to be
{\it first order} where $\Psi$ is infinite, irrespective of
$\sigma_0$. The large but finite $\Psi$ values found in experiment
could be attributed to {\it finite size} rounding effects of a
{\it first order} transition rather than to a very small
$\sigma_0$.

We believe that the vast amount of melting curves data existing in
the literature should be reevaluated using the correct exponent
$c$ in order to obtain a realistic estimate for the cooperativity
parameter. Certainly this should result in a larger cooperativity
parameter $\sigma_0$ which would in turn be reflected in an even
smaller persistence length of the stick-and-joint structure of the
molecule. It would also be very interesting to measure $\sigma_0$
directly from single molecule experiments using, for example,
fluorescence correlation spectroscopy.

We thank H. A. Scheraga and E. Yeramian for helpful discussions.

\noindent Y. Kafri and D. Mukamel

Department of Physics of Complex Systems, The Weizmann Institute
of Science, Rehovot 76100, Israel

\noindent L. Peliti

Dipartimento di Scienze Fisiche and Unit\`a INFM, Universit\`a
``Federico II'', Complesso Monte S. Angelo, I--80126 Napoli, Italy

\end{document}